\documentclass{article}
\usepackage[english]{babel}

\usepackage[utf8]{inputenc}
\RequirePackage{amsthm,amsmath,amsfonts,amssymb}

\usepackage{adjustbox}
\usepackage{algorithm}

\usepackage{algpseudocode}
\usepackage{array}

\usepackage[english]{babel}
\usepackage{bigints}
\usepackage{booktabs}

\usepackage{caption}
\usepackage{enumerate}
\usepackage{graphicx,psfrag,epsf}
\usepackage{soul}
\usepackage{listings}
\usepackage{lscape}
\usepackage{multirow}
\usepackage{makecell}

\usepackage{rotating}
\usepackage[rightcaption]{sidecap}
\usepackage[flushleft]{threeparttable}
\usepackage{wrapfig}
\usepackage[table,xcdraw]{xcolor}

\definecolor{imsblue}{RGB}{0,0,100}
\usepackage[colorlinks]{hyperref}
\hypersetup{
    colorlinks=true,
    linkcolor=blue,
    urlcolor=blue,
    citecolor=blue
}

\newtheorem{innercustomgeneric}{\customgenericname}
\providecommand{\customgenericname}{}
\newcommand{\newcustomtheorem}[2]{%
  \newenvironment{#1}[1]
  {%
   \renewcommand\customgenericname{#2}%
   \renewcommand\theinnercustomgeneric{##1}%
   \innercustomgeneric
  }
  {\endinnercustomgeneric}
}

\newcustomtheorem{customthm}{Theorem}
\newcustomtheorem{customlemma}{Lemma}

\providecommand{\keywords}[1]
{
  \small	
  \textbf{\textit{Keywords:}} #1
}

\title{\Large Multiple Imputation Method for High-Dimensional Neuroimaging Data}
\date{}
\author{\small Tong Lu$^{1}$,  Chixiang Chen$^{2}$, Hsin-Hsiung Huang$^{3}$, Peter Kochunov$^{4}$, Elliot Hong$^{4}$, Shuo Chen$^{2,5}$\\
\footnotesize
$^{1}$Department of Mathematics, University of Maryland,
College Park \\
\footnotesize
$^{2}$Division of Biostatistics and Bioinformatics, School of Medicine, University of Maryland \\
\footnotesize
$^{3}$Department of Statistics and Data Science, University of Central Florida \\
\footnotesize
$^{4}$Department of Psychiatry and Behavioral Science,  University of Texas Health Science Center \\
\footnotesize
$^{5}$Maryland Psychiatric Research Center, School of Medicine, University of Maryland \\
\footnotesize *\href{mailto:shuochen@som.umaryland.edu}{shuochen@som.umaryland.edu}
}
\vspace{-3mm}

\begin{document}
\maketitle
\begin{sloppypar}

\begin{abstract}
Missingness is a common issue for neuroimaging data, and neglecting it in downstream statistical analysis can introduce bias and lead to misguided inferential conclusions. It is therefore crucial to conduct appropriate statistical methods to address this issue.  
While multiple imputation is a popular technique for handling missing data, its application to neuroimaging data is hindered by high dimensionality and complex dependence structures of multivariate neuroimaging variables. 
To tackle this challenge, we propose a novel approach, named \textbf{H}igh d\textbf{i}mensional \textbf{M}ultiple Imput\textbf{a}tion (HIMA), based on Bayesian models. HIMA develops a new computational strategy for sampling large covariance matrices based on a robustly estimated posterior mode, which drastically enhances computational efficiency and numerical stability.  
To assess the effectiveness of HIMA, we conducted extensive simulation studies and real-data analysis using neuroimaging data from a Schizophrenia study. HIMA showcases a computational efficiency improvement of over 2000 times when compared to traditional approaches, while also producing imputed datasets with improved precision and stability.
\end{abstract}


\section{Introduction}
\label{HIMA_Intro}

\begin{figure}[h]
\makebox[\textwidth][c]{
    \includegraphics[width=0.8\textwidth]{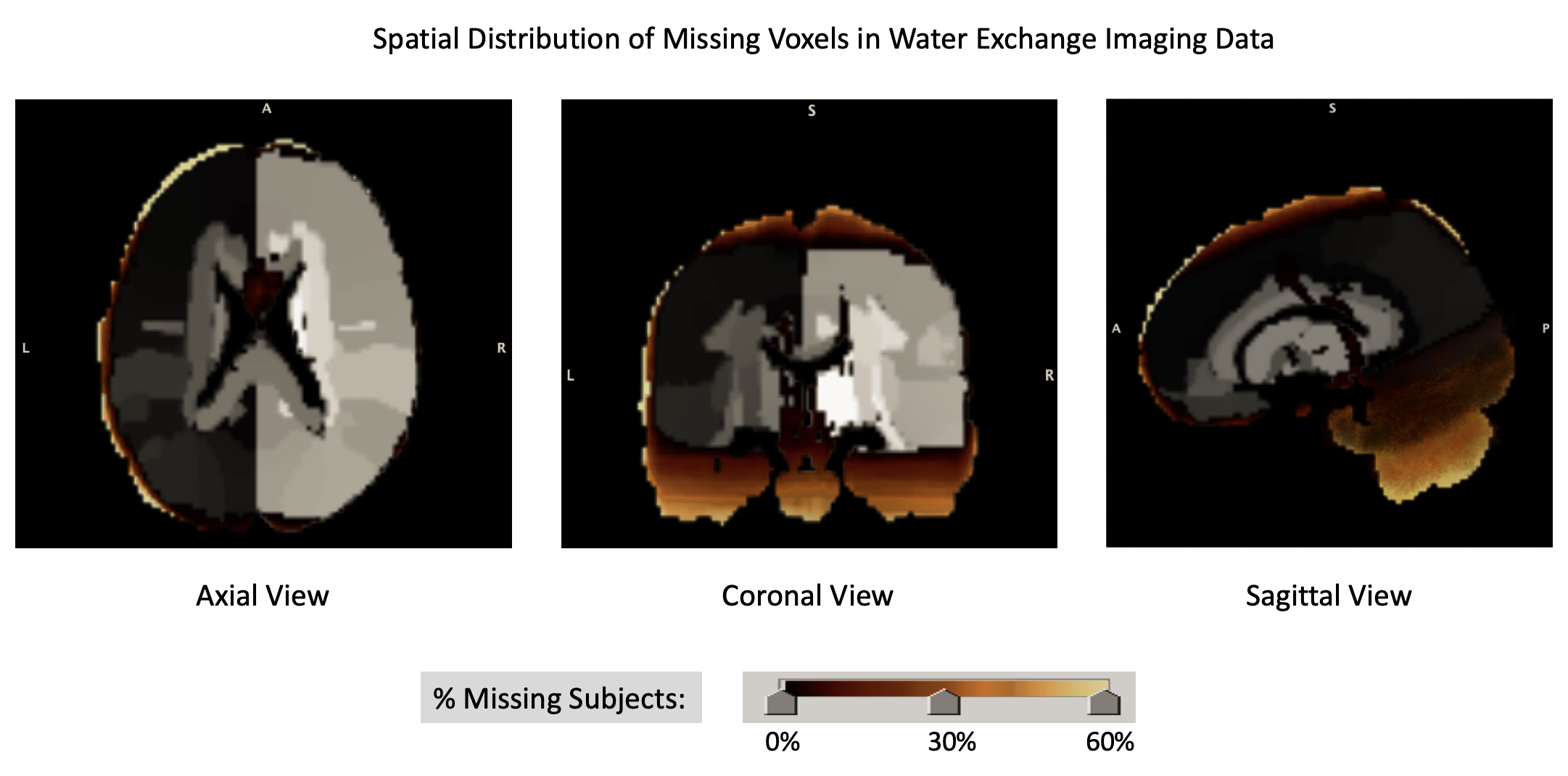}
}
\caption[An illustration of missingness distribution in neuroimaging data]{\footnotesize \textit{An illustration of missingness distribution in a neuroimaging dataset. Neurovascular water exchange imaging data were collected from a schizophrenia research study ($N=58$). The visualization presents different cross-sectional views, with grey voxels representing non-missing data and the gold color scale indicating the percentage of subjects with missing data. Darker-colored voxels contain missing data from a smaller proportion of subjects and vice versa.}}
\label{tab:HIMA_spatial_missing}
\end{figure}

Neuroimaging data are fundamental for studying the brain's structure and function, providing valuable insights into various neurological disorders and cognitive processes. Missing data, however, occur frequently in brain imaging research due to limited image acquisition and susceptibility artifacts, causing signal loss and spatial distortion in the images \cite{mulugeta2017methods}. An example of the spatial distribution of missing voxels in a magnetic resonance imaging (MRI) dataset is shown in \autoref{tab:HIMA_spatial_missing}.
Despite advancements in statistical techniques for processing imaging data, the proper handling of neuroimaging missingness remains inadequately studied, which impedes the accurate analysis and interpretation of findings. For example, missing data can lead to biased estimation, reduce statistical power, and limit the generalizability of results \cite{baraldi2010introduction, vaden2012multiple,newman2014missing}. To address these challenges, we are motivated to propose a practical yet robust multivariate multiple imputation technique specifically designed for high-dimensional neuroimaging data. 



Nowadays, commonly used strategies to handle incomplete data include (i) complete data analysis, (ii) single imputation, and (iii) multiple imputation (MI). 
In certain scenarios, particularly when the missingness is minimal (e.g.,  less than 5$\%$)  and occurs completely at random, complete case analysis can be considered the best approach to prevent analysis bias. However, neuroimaging data often exhibit complex missing patterns that do not conform to such straightforward criteria. 
Additionally, simply omitting these voxels may risk excluding brain regions of particular research interest and may be costly to imaging spatial coverage, especially along cortical boundaries.  This, in turn, could raise the risk of Type II errors.
Improving upon (i), simple imputation involves replacing individual missing value with a single value, often using methods like mean or mode substitution. While simple imputation may offer quick solutions in certain situations, it frequently introduces bias into the data (e.g., artificial reductions in variability) and results in overly precise results without accounting for any uncertainty.
\cite{schafer1997analysis} and \cite{rubin2004multiple} addressed this by developing MI techniques that can incorporate uncertainty about the unknown missing values. MI replaces each missing value with a set of plausible values imputed based on two factors: (a) the observed values for a given subject; (b) the relationships observed in the data for other subjects. 

\begin{figure}[h]
    \makebox[\textwidth][c]{
        \includegraphics[width=0.4\textwidth]{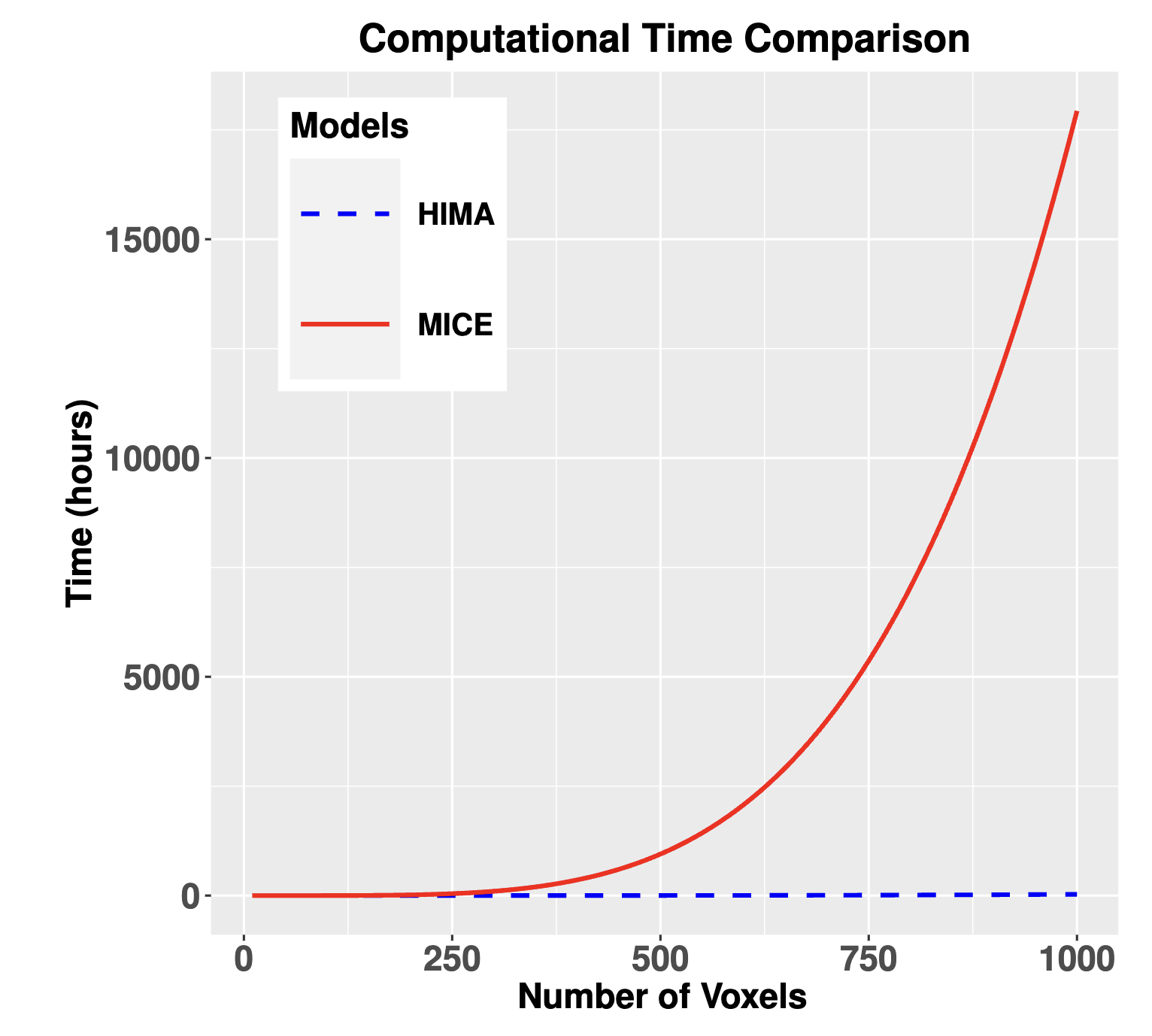}
    }
    \caption[Running time against the number of voxels using MICE and HIMA]{\footnotesize \textit{ The running time (in hours) is plotted against the number of voxels for both MICE and HIMA (the proposed model). We set the number of imputed datasets to be 10; within each dataset, there are 15 iterations. Both MICE and HIMA were implemented using a single computational node with 128 GB memory and an 8-core CPU. }}
    \label{tab:HIMA_time_comp}
\end{figure}

Statistical literature on MI techniques has surged \cite{Murray2018, carpenter2023multiple}. Applying MI to neuroimaging data, however, is limited, primarily due to computational tractability issues. Take \textbf{M}ultivariate \textbf{I}mputation by \textbf{C}hained \textbf{E}quations (MICE), one of the most commonly used MI toolboxes, as an example \cite{royston2011multiple,vaden2012multiple,enders2022applied}. Given approximate normal data following $\mathcal{N}(\boldsymbol{\mu},\boldsymbol{\Sigma})$, MICE specifies an inverse Wishart distribution as a conjugate prior distribution for the covariance $\mathbf{\Sigma}$.
Sampling a large $\boldsymbol{\Sigma}_{p\times p}$ matrix (e.g., $p=1000$, as is common in neuroimaging data) from the corresponding posterior inverse Wishart distribution 
can become computationally unstable and intractable. This may lead to inaccuracies in subsequent data sampling, which ultimately impacts the overall precision of imputation results.
Additionally, sampling $\boldsymbol{\Sigma}_{p\times p}$ 
involves matrix inversion, which requires a computational cost of $\mathcal{O}(p^3)$. This cubic time complexity will be further compounded by the number of sampling iterations and the total number of imputed datasets needed.
In \autoref{tab:HIMA_time_comp}, we have shown the time required to impute missingness in a real MRI dataset from a schizophrenia study using the MICE package. Notably, computational time increases exponentially as the number of voxels grows. To impute missingness in a typical brain region with hundreds of variables, it can take thousands of hours to run MICE, which is not so computationally feasible.

    
Computational complexity and tractability can pose significant bottlenecks for handling high-dimensional imaging data, as indicated by \autoref{tab:HIMA_time_comp}. To address this challenge, we propose a new \textbf{H}igh-\textbf{D}imensional \textbf{M}ultiple \textbf{I}mput\textbf{a}tion (HIMA) method, designed specifically for high-dimensional neuroimaging data. HIMA adopts the commonly used Bayesian framework through Markov chain Monte Carlo (MCMC): 
it first implements the same imputation step as in classical MI techniques to impute missing entries by considering a joint multivariate normal model; next, it modifies the posterior step by updating the normal covariance matrix with a robustly estimated posterior mode. The posterior mode represents the most probable draws from the posterior distribution of covariance, which is essentially identical to the maximum likelihood estimates of the likelihood functions \cite{schafer1997analysis}.  
We propose a posterior mode estimator that is suitable for situations where $n\ll p$ and have established  asymptotic properties of it.


This article presents three main contributions. Firstly, we introduce a novel MI technique called HIMA, specifically tailored for high-dimensional neuroimaging data. HIMA substantially alleviates the computational burden from $\mathcal{O}(p^3)$ to $\mathcal{O}(p)$ per MCMC iteration. 
Secondly, extensive simulation studies demonstrate reduced bias and dispersion in the imputed data generated by HIMA. The imputation expands brain map coverage, which in turn improves the interpretation of imaging results.
Lastly, we have developed a user-friendly package for implementing HIMA, making it easily accessible and convenient for researchers to apply in their neuroimaging studies.

The rest of this paper is structured as follows. In Section 2, we introduce the HIMA method, posterior mode estimation, and imputation algorithms. In Section 3, we assess the performance of HIMA using both semi-synthetic and real MRI imaging datasets and comparing it to frequently used imputation methods. We conclude with a discussion in Section 4.

\section{Methods}

\subsection{Background}


Our proposed imputation method HIMA is designed for voxel-level neuroimaging data, such as voxel-level hemodynamic response for fMRI\footnote{Functional Magnetic Resonance Imaging}, voxel-level fractional anisotropy for DTI\footnote{Diffusion Tensor Imaging}, and ALFF\footnote{Amplitude of low-frequency fluctuation} for rs-fMRI\footnote{Resting state fMRI}. HIMA can be easily adaptable to region-level data as well, offering versatility across various neuroimaging applications.

Without loss of generality, we let $\boldsymbol{y}_i= \{y_{ij}\}_{j =\{1,\dots, p\}}$ denote the brain signals of interest for the $i$-th subject with $i =\{1,\dots, n\}$, where $j$ represents the $j$-th voxel. Typically, brain signals exhibit approximate normal distribution characteristics \cite{chavez2010functional,maier2004biexponential}; we thus consider a joint multivariate normal (MVN) model for $\boldsymbol{y}_i$. 
Specifically,  we express the observed and missing part of $\boldsymbol{y}_i$ by $\boldsymbol{y}_i^\text{obs}$ and $\boldsymbol{y}_i^\text{mis}$, and assume that 
they follow:
\begin{align}
\label{HIMA_y_dist}
\begin{pmatrix}\boldsymbol{y}_i^\text{obs}\\
    \boldsymbol{y}_i^\text{mis}
\end{pmatrix} &\sim  \mathcal{N}
\begin{bmatrix}
\begin{pmatrix}
\boldsymbol{\mu}_\text{obs}\\
\boldsymbol{\mu}_\text{mis}
\end{pmatrix}\!\!,&
\left(\begin{array}{cc} \mathbf{\Sigma}_\text{obs,obs} & \mathbf{\Sigma}_\text{obs,mis}\\ \mathbf{\Sigma}_\text{mis,obs} & \mathbf{\Sigma}_\text{mis,mis} \end{array}\right)
\end{bmatrix},
\end{align} 
where 
$\begin{pmatrix}
    \boldsymbol{\mu}_\text{obs}\\
    \boldsymbol{\mu}_\text{mis}
\end{pmatrix}$ is the partitioned mean vector and the four sub-covariance-matrices are partitioned from covariance $\mathbf{\Sigma}$. $\mathbf{\Sigma}$ is crucial for jointly leveraging different levels of associations, including voxel-wise and subject-wise associations, during the imputation process.
Under the assumption of \textit{missing at random} (commonly adopted for imputing neuroimaging data \cite{bartlett2015multiple,mulugeta2017methods}), 
our goal is to impute $\boldsymbol{Y}_\text{mis}=\{\boldsymbol{y}_i^\text{mis}\}_{i=1}^n$ based on the observed data $\boldsymbol{Y}_\text{obs}=\{\boldsymbol{y}_i^\text{obs}\}_{i=1}^n$  while preserving the uncertainty of $\boldsymbol{Y}_\text{mis}$.

Bayesian models are widely used for handling multivariate MI applications \cite{royston2011multiple, little2002bayes}. In Bayesian models, parameters (e.g., $\boldsymbol{\mu}$ and $\boldsymbol{\Sigma}$) and $\boldsymbol{Y}_\text{mis}$ can be iteratively updated until convergence, where the imputed datasets can be sampled from converged posterior distribution.  
Nonetheless, computational challenges arise in the classic Bayesian-based MI methods, especially for data with high dimensions \cite{khan2020sice}). For example, 
when imputing a missing dataset of 500 variables using a single node, MICE, a classical MI based method, may take up to 2300 hours (see Figure \ref{tab:HIMA_time_comp}). 

\subsection{HIMA method}
To address the aforementioned challenges, we propose HIMA, a relaxed multivariate imputation approach designed for handling missingness in data with high dimension ($n \ll p$). HIMA follows established data augmentation algorithms \cite{tanner1987calculation,li1988imputation, gelman2014bayesian} and  MCMC procedures for multivariate data imputation. 
Specifically, HIMA iteratively imputes $\boldsymbol{Y}^{[t]}_{\text{mis}}$ ($t=1, \cdots, T$ is the iteration index), and updates $\boldsymbol{\mu}^{[t]}$ and $\boldsymbol{\Sigma}^{[t]}$   until convergence. The detailed iteration steps are provided below:

\vspace{1mm}
\textbf{\textit{1. I-step (Impute }}$\boldsymbol{Y}^{[t]}_{\text{mis}}$). Given parameters $\boldsymbol{\mu}^{[t-1]}$  and $\boldsymbol{\Sigma}^{[t-1]}$ at the $(t-1)$-th iteration, we generate MVN missing values $\boldsymbol{Y}^{[t]}_{\text{mis}}$ by
\begin{align}
    \boldsymbol{Y}^{[t]}_{\text{mis}} \sim \mathcal{N}_{\boldsymbol{Y}_{\text{mis}} | \boldsymbol{Y}_{\text{obs}}}(\boldsymbol{\mu}^{[t-1]}_{\text{mis}|\text{obs}},\boldsymbol{\Sigma}^{[t-1]}_{\text{mis}|\text{obs}}),
\end{align}
where \begin{align*}
    &\boldsymbol{\mu}_{\text{mis}|\text{obs}}=
\boldsymbol{\mu}_{\text{mis}} +
                    \mathbf{\Sigma}_{\text{mis},\text{obs}}
    {\mathbf{\Sigma}_{\text{obs},\text{obs}}}^{-1} 
        (\boldsymbol{Y}_{\text{obs}}-\boldsymbol{\mu}_{\text{obs}}),\\
&\mathbf{\Sigma}_{\text{mis}|\text{obs}}=\mathbf{\Sigma}_{\text{mis},\text{mis}}-\mathbf{\Sigma}_{\text{mis},\text{obs}} {\mathbf{\Sigma}_{\text{obs},\text{obs}}}^{-1}\mathbf{\Sigma}_{\text{obs},\text{mis}}.
\end{align*}


\textbf{\textit{2. P-step (Update $\boldsymbol{\Sigma}^{[t]}$ and $\boldsymbol{\mu}^{[t]}$)}}. 
After imputing $\boldsymbol{Y}^{[t]}_{\text{mis}}$ and augmenting it with $\boldsymbol{Y}_{\text{obs}}$, we proceed to update parameters $\boldsymbol{\Sigma}^{[t]}$ and $\boldsymbol{\mu}^{[t]}$ using a standard MCMC procedure. We introduce a relaxed parameter sampling strategy in the MCMC procedure outlined as follows:

\textbf{\textit{Update $\boldsymbol{\Sigma}^{[t]}$.}}
The traditional approach for sampling $\boldsymbol{\Sigma}$ relies on a posterior distribution with an inverse Wishart conjugate prior distribution: $\boldsymbol{\Sigma} \sim \textbf{W}^{-1}(\boldsymbol{\Psi},\nu)$, where $\boldsymbol{\Psi}$ is a positive definite scale matrix and $\nu$ is a degrees of freedom.
Accordingly, the posterior distribution becomes $\textbf{W}^{-1}(\boldsymbol{\Psi}+n\boldsymbol{S},\nu+n)$, where $\boldsymbol{S}$ is the sample covariance.
In practice, sampling $\boldsymbol{\Sigma}^{[t]}$ with a high-dimensional $p$ is challenging due to the computational intractability and instability \cite{bandiera2010knowledge, diakonikolas2019robust}.  A sound remedy for sampling large posterior $\boldsymbol{\Sigma}$ is to estimate its \textit{Maximum a Posterior} (MAP), the value that is most likely to be sampled (i.e., the mode) \cite{van2011optimal, wang2018randomized}. Specifically, given a complete sample augmented by previous imputed $\boldsymbol{Y}^{[t]}_{\text{mis}}$, we estimate the posterior mode by 
\begin{align}
\label{max_sig}
        \boldsymbol{\Sigma}^{[t]}
        &= \arg \max_{\mathbf{\hat{\Sigma}}} ~p(\mathbf{\Sigma}|\boldsymbol{Y}_\text{obs}, \boldsymbol{Y}^{[t]}_{\text{mis}})    \nonumber \\
        &= \arg \max_{\mathbf{\hat{\Sigma}}}  \textbf{W}^{-1}(\boldsymbol{\Psi}+n\boldsymbol{S},\nu+n) \\
        &=\arg \max_{\mathbf{\hat{\Sigma}}} \Big( \frac{|\boldsymbol{\Psi}+n\boldsymbol{S}|^{\frac{\nu+n}{2}}}{2^{\frac{(\nu+n)p}{2} } \Gamma_p\left(\frac{\nu+n}{2}\right)}|\mathbf{\Sigma}|^{\frac{-(\nu+n+p+1)}{2} } \text{exp} \big(-\frac{1}{2} \operatorname{tr}( (\boldsymbol{\Psi}+n\boldsymbol{S}) \mathbf{\Sigma}^{-1})\big) \Big).
        \nonumber
\end{align}
It is generally challenging to solve \autoref{max_sig} directly. We develop a new computational approach to implement the optimization step introduced in Section \ref{sec_Implementation}. 

\vspace{2mm}

\textbf{\textit{ Update $\boldsymbol{\mu}^{[t]}$.}}
Given the augmented data $[\boldsymbol{Y}_\text{obs}, \boldsymbol{Y}^{[t]}_{\text{mis}}]$ and updated $\boldsymbol{\Sigma}^{[t]}$, we generate the posterior mean $\boldsymbol{\mu}^{[t]}$ by
\begin{align}
        \boldsymbol{\mu}^{[t]} | \boldsymbol{Y}_\text{obs}, \boldsymbol{Y}^{[t]}_{\text{mis}}, \boldsymbol{\Sigma}^{[t]} \sim \mathcal{N}_{\boldsymbol{\mu}}(\boldsymbol{\bar{Y}},\frac{\boldsymbol{\Sigma}^{[t]}}{n}),
\end{align}
where $\bar{Y}$ is the mean vector of the augmented data $[\boldsymbol{Y}_\text{obs}, \boldsymbol{Y}^{[t]}_{\text{mis}}]$. The posterior mean is derived based on a non-informative prior for $\boldsymbol{\mu}$ (uniform over the $p$-dimensional real space) \cite{schafer1997analysis}.

\textbf{\textit{Remarks on updating $\boldsymbol{\Sigma}^{[t]}$.}}
In Bayesian analysis, 
MAP estimation
is designed to maximize a conditional probability distribution. Sampling the mode of a posterior distribution has been carefully studied in the statistical literature, with notable examples including \cite{geman1984stochastic,gelman1995bayesian,doucet2002marginal}. As pointed out in \cite{schafer1997analysis}, maximizing $ p(\mathbf{\Sigma}| \boldsymbol{Y})$ is nearly identical to obtaining the ML estimates by maximizing the normal likelihood $L(\mathbf{\Sigma}| \boldsymbol{Y})= \Pi_{i=1}^n ~p(\textbf{y}_i|\mathbf{\Sigma})$. 
Additionally, updating $\mathbf{\Sigma}^\text{mode}$ eliminates the need to compute the inverse of a large covariance matrix $\mathbf{\Sigma}_{p \times p}$. This relaxation reduces the computational burdens from $\mathcal{O}(Cp^3)$ to $\mathcal{O}(Cp)$, where $C$ depends on factors including the number of observations, iterations in MCMC and the number of imputed datasets. Updating $\mathbf{\Sigma}^\text{mode}$ also avoids the sampling process of large matrices at each individual iteration, which enhances computational traceability and ultimately the imputation results.
Lastly, the posterior mode can be estimated by empirical Bayesian methods. In the following section, we introduced a tailored approach particularly designed for data with $n\ll p$.


\subsubsection{Estimating posterior mode $\mathbf{\Sigma}$}
\label{sec_Implementation}
In this section, we present a new computational strategy to estimate the posterior mode of $\mathbf{\Sigma}$ following \autoref{max_sig} under the scenario of $n\ll p$.
Following many prior works on posterior mode estimation of $\mathbf{\Sigma}$, we adopt an unorthodox representation for inverse Wishart distribution, denoting as $\textbf{W}^{-1}(\boldsymbol{\boldsymbol{\zeta}},\lambda)$, to facilitate easier demonstration \cite{champion2003empirical}. Here, $\boldsymbol{\zeta}$ represents the mean and $\lambda$ represents a measure of precision depending on the standard degrees of freedom $v$ by $v=\lambda+p+1$. With these notations in place, the prior density can be written as
\begin{align*}
    \boldsymbol{\zeta} \sim \frac{|\lambda \boldsymbol{z}|^{\frac{\lambda+p+1}{2}}}{2^{\frac{p(\lambda+p+1)}{2}} \pi^{\frac{p(p-1)}{4}} \prod_{j=0}^{p-1} \Gamma\left(\frac{\lambda+j}{2}+1\right)}|\boldsymbol{\zeta}|^{\frac{-(\lambda+2 p+2)}{2}} \exp \left\{-\frac{1}{2} \operatorname{tr} (\lambda \boldsymbol{z} \boldsymbol{\zeta}^{-1})\right\},
\end{align*}
where $\lambda \boldsymbol{z}=\boldsymbol{\Psi}>0$ ($\boldsymbol{\Psi}$ is the scale matrix in the traditional notation for inverse Wishart distribution). 
Accordingly, we update the posterior distribution of $\mathbf{\Sigma}$ proportional to
\begin{align*}
    |\boldsymbol{\zeta}|^{\frac{-(n+\lambda+2 p+2)}{2}} \exp \left\{-\frac{1}{2} \operatorname{tr} (\lambda \boldsymbol{z}+S) \boldsymbol{\zeta}^{-1}\right\},
\end{align*}
and further derive the posterior mode as
\begin{align}
\label{mode_est}
\mathbf{\Sigma}^{\text{mode}}=\frac{\lambda\boldsymbol{z}+\boldsymbol{S}}{n+\lambda+2 p+2},
\end{align}
where $\boldsymbol{S}$ is the sample covariance 
and $\lambda$, $\boldsymbol{z}$ are two parameters to be estimated. 
We adopt an empirical Bayesian approach proposed in \cite{champion2003empirical} to estimate $\lambda$ and $\boldsymbol{z}$. Specifically, the estimation criterion is to minimize the expected estimation loss between $\hat{\mathbf{\Sigma}}^{\text{mode}}(\hat{\lambda}, \hat{\boldsymbol{z}})$ (posterior mode estimator) and $\mathbf{\Sigma}_0$ (true parameter) using the Kullback–Leibler distance:
\begin{align}
\label{HIMA_est_loss}
\arg \min_{\hat{\lambda}, \hat{\boldsymbol{z}}}
KL\Big( \mathbf{\Sigma}_0,\hat{\mathbf{\Sigma}}^{\text{mode}}(\hat{\lambda}, \hat{\boldsymbol{z}}) \Big)= \arg \min_{\hat{\lambda}, \hat{\boldsymbol{z}}}
    \int  p(\boldsymbol{Y} \mid \mathbf{\Sigma}_0) \log \frac{p(\boldsymbol{Y} \mid \mathbf{\Sigma}_0)}{p\Big( \boldsymbol{Y} \mid \hat{\mathbf{\Sigma}}^{\text{mode}}(\hat{\lambda}, \hat{\boldsymbol{z}}) \Big)}\mathrm{d} \boldsymbol{Y}.
\end{align}
The detailed estimation procedure and the empirically based estimates $\hat{\lambda}, \hat{\boldsymbol{z}}$ are provided in \textbf{Appendix A}, where no steps in the procedure require $n \gg p$, which is desirable in our application. 
We further establish the theoretical guarantee that  $\mathbf{\hat{\mathbf{\Sigma}}^{\text{mode}}}(\hat{\lambda}, \hat{\boldsymbol{z}})\stackrel{n}{\longrightarrow} \mathbf{\Sigma}_0$, where $\mathbf{\Sigma}_0$ denotes the true covariance parameter (see details in \textbf{Appendix B.1}). Therefore, the mode covariance estimated by (\ref{HIMA_est_loss}) asymptotically converges to the true parameter. 

\subsubsection{Algorithm}
We implement HIMA by Algorithm~\ref{alg:HIMA_alg1}. In the algorithm, $\boldsymbol{Y}_{(0)}$ denotes the initial dataset. Since HIMA is a MI procedure, we impute $\boldsymbol{Y}_{(0)}$ for $M$ times by sampling from the posterior distribution and denote the  $m$-th imputed dataset as  $\boldsymbol{Y}_{(m)}$ ($m\in[M]$). 
For each $\boldsymbol{Y}_{(m)}$, we iteratively sample the multivariate missing data $\boldsymbol{Y}_{\text{mis}}$ and the parameters $\boldsymbol{\mu}$ and $\boldsymbol{\Sigma}$ by following the two steps in the previous section. The posterior distributions converge at the $T$-th iteration and we now obtain one imputed dataset $\boldsymbol{Y}_{(m)}$.
The output of the algorithm is $M$ imputed datasets $\{\boldsymbol{Y}_{(m)}\}_{m\in M}$. Subsequently, statistical inference can be further made by collectively pooling the estimated results by various pooling methods (e.g., Rubin’s rules outlined in \cite{rubin2004multiple}).

\begin{algorithm}
    \caption{HIMA}
    \label{alg:HIMA_alg1}
    \begin{algorithmic}[1]
    \Procedure{Algorithm}{Input: $\boldsymbol{Y}_{(0)}, M, T$}
        
        
        
       
       \For {$m=1,2,\ldots, M$ imputed datasets~} 
           \State Impute $\boldsymbol{Y}_{(0)}$ initially\footnotemark
    
            \State Initialize $\boldsymbol{Y}_{(m)}=\boldsymbol{Y}_{(0)}$,  $\boldsymbol{\mu}^{[0]}=\boldsymbol{\bar{Y}}_{(0)}$ (\text{sample mean}), $\boldsymbol{\Sigma}^{[0]}=\boldsymbol{S}_{\boldsymbol{Y}_{(0)}}$ (\text{sample covariance})
            
            \For {$t=1,2,\ldots,T$ iterations~}
                \For {$i=1,2,\ldots,n$ subjects~}

        \State Impute $\boldsymbol{Y}^{[t+1,i]}_{\text{mis}} | \boldsymbol{Y}_{\text{obs}} \sim \mathcal{N} (\boldsymbol{\mu}^{[t]}_{\text{mis}|\text{obs}},\mathbf{\Sigma}^{[t]}_{\text{mis}|\text{obs}})$\\
                \State ~~~~~~ where $\boldsymbol{\mu}^{[t]}_{\text{mis}|\text{obs}} =
                    \boldsymbol{\mu}^{[t]}_{\text{mis}} +
                    \mathbf{\Sigma}_{\text{mis},\text{obs}}^{[t]}
        {\mathbf{\Sigma}_{\text{obs},\text{obs}}^{[t]}}^{-1} 
        (\boldsymbol{Y}^{[t,i]}_{\text{obs}}-\boldsymbol{\mu}^{[t]}_{\text{obs}})$,
        
        \\ \State ~~~~~~~~~~~~~~$\mathbf{\Sigma}^{[t]}_{\text{mis}|\text{obs}}=\mathbf{\Sigma}^{[t]}_{\text{mis},\text{mis}}-\mathbf{\Sigma}^{[t]}_{\text{mis},\text{obs}} {\mathbf{\Sigma}^{[t]}_{\text{obs},\text{obs}}}^{-1}\mathbf{\Sigma}^{[t]}_{\text{obs},\text{mis}}$

                    \State Estimate posterior mode $\mathbf{\Sigma}^{[t]} |(\boldsymbol{Y}_{\text{obs}} ,\boldsymbol{Y}^{[t,i]}_{\text{mis}|\text{obs}},\boldsymbol{\mu}^{[t]})$ by (\ref{HIMA_est_loss})
                    \State Sample $\boldsymbol{\mu}^{[t]} \sim p(\boldsymbol{\mu} |\mathbf{\Sigma}^{[t]},\boldsymbol{Y}_{\text{obs}} ,\boldsymbol{Y}^{[t,i]}_{\text{mis}} )$

        
                \EndFor
                \State Fill in $\boldsymbol{Y}^{[t]}_{\text{mis}}=\{\boldsymbol{Y}^{[t,i]}_{\text{mis}}\}_{i=1}^n$ for all $n$ subjects
            \EndFor

            \State Obtain the $m$-th fully imputed dataset $\mathbf {Y}_{(m)}=[\boldsymbol{Y}_{\text{obs}} ,\boldsymbol{Y}^{[t]}_{\text{mis}} ]$ 
            
        \EndFor
        \State \textbf{return} $\{\boldsymbol{Y}_{(m)}\}_{m\in[M]}$
    \EndProcedure
    \end{algorithmic}
\end{algorithm}
\footnotetext{Using feasible method (mean imputation, median imputation, hot deck imputation, etc.)}


To summarize, HIMA retains the traditional \textit{imputation I-step} in classical MI and introduces relaxation to the \textit{posterior P-step} to accommodate the large covariance matrix.
Both steps are iterated sufficiently long until the sequence
$\{(\boldsymbol{Y}^{[t]}_\text{mis},\boldsymbol{\mu}^{[t]},\mathbf{\Sigma}^{[t]}): t=1,2,\dots\}$ 
(an MCMC sequence) converges to a stationary distribution $\mathbb{P}(\boldsymbol{Y}_\text{mis},\boldsymbol{\mu},\mathbf{\Sigma}| \boldsymbol{Y}_\text{obs})$.
We justify Algorithm~\ref{alg:HIMA_alg1} from a rigorous Bayesian point of view in \textbf{Appendix B.2}.
Simulations have indicated that HIMA can perform well with a small number of imputations $M$ (e.g., $5-20$). The number of iterations $T$ can be defined by users. In practice, a few cycles (e.g., $20-30$) is typically sufficient to ensure the convergence of the distributions of parameters and imputed values. 

\section{Results}
A real MRI dataset from a schizophrenia study \cite{goldwaser2023evidence} is used to assess the accuracy and efficiency of HIMA in comparison to existing imputation methods. The study focuses on cross-sectional neurovascular water exchange (Kw) data collected from 58 subjects (age: $37.8 \pm 14.0$; sex, 37 M: 21 F) with schizophrenia spectrum disorder. HIMA is applied to two different versions of the data:
(i) semi-synthetic data, which includes originally unknown missing entries and known missing entries that are artificially  inserted;
(ii) real data, which contains completely unknown missing values.

The subjects were recruited from mental health clinics and through media advertisements in Baltimore, MD. Imaging data were collected using a diffusion-prepared arterial spin labeling protocol from 2019 to 2022 and preprocessed using ASL (BASIL) pipeline \cite{shao2019mapping, zhang2022improving, goldwaser2023evidence}. 
Detailed information on the imaging acquisition and preprocessing procedures can be found in \textbf{Appendix C}. This schizophrenia study aims to investigate the association between neural-capillary water exchange function and schizophrenia spectrum disorder, with a specific focus on 99 brain regions of interest obtained using the International Consortium for Brain Mapping (ICBM) brain template.

\subsection{Semi-synthetic data analysis}
The primary measure of this schizophrenia study was the whole-brain average Kw values. We intend to apply our method to impute voxel-level missing values within each of the 99 regions individually because within-region Kw values are typically more homogeneous and strongly associated. Accordingly, the data structure to be imputed is denoted as ${y_{ij}}^{(r)}$, where $i\in[58]$ represents subjects, $r\in[99]$ represents regions, and the number of voxels $j\in[p^{(r)}]$ depends on region $r$

The overall missing rate of the data ${{y_{ij}}^{(r)}: r\in[99]}$ is $14.60\%$. 
We construct a \textit{semi-synthetic} dataset by artificially removing voxel values randomly $(i, j)$, i.e., random voxel locations of random subjects. Specifically, we randomly remove $t$ data entries from each voxel vector, where $t$ ranges from 1 to 8. On average, this process results in an additional 4.5 missing spots (out of 58) in each voxel vector.
The missing rate of the semi-synthetic dataset is now $20.34\%$, compared to the original rate of $14.60\%$. 

We apply several processing steps to the semi-synthetic dataset as follows. First, we drop out voxel vectors with remarkably high missing rates, setting the threshold at $40\%$ to ensure robust performance. Experiments demonstrate that in this dataset, HIMA may become unstable and less accurate if the voxel-vector-level threshold exceeds $40\%$. On average, only $3.17\%$ of voxels exceed the threshold within each region. For example, in a region with 800 voxels, only 25 voxels need to be dropped out.
Next, we perform kernel smoothing on subjects and align their smoothed probability density estimates to account for subject-level random effects. 
Lastly, we perform HIMA to impute the missing data on each post-processed ${\boldsymbol{Y}^{(r)}}^*=\{y^*_{{ij}^{(r)}}\}$, where we set $M=15$ (number of imputed datasets) and $T=20$ (number of iterations).  

Since these missing voxels are artificially removed, we can compare the imputed values with `true' (but removed) values to assess the performance of imputation algorithms.
We first define an $n\times p$ indicator matrix $\boldsymbol{\tau}$ for a data matrix $\boldsymbol{Y}=\{y_{ij}\}_{i\in[n],j\in[p]}$, where each element $\tau_{ij}=1$ if $y_{ij}$ is observed; $\tau_{ij}=0$ otherwise. Using $\boldsymbol{\tau}$, we assess imputation accuracy by the following metrics: 

(i) Weighted mean absolute error (wMAE): 
\begin{align*}
    \frac{ \sum_{i\in[n]} \sum_{j\in[p]} I(\tau_{ij}=0)\times |y_{ij}^\text{imputed} - y_{ij}^\text{true}|~ / \sqrt{\text{Var}(Y_{.j})} }{\sum_{i\in[n]} \sum_{j\in[p]} I(\tau_{ij}=0)}
\end{align*}
for a single imputed dataset. We then compute the mean and standard deviation of wMAE across all $M$ imputed datasets. 

(ii) Weighted mean square error (wMSE): 
\begin{align*}
    \frac{ \sum_{i\in[n]} \sum_{j\in[p]} I(\tau_{ij}=0)\times (y_{ij}^\text{imputed} - y_{ij}^\text{true})^2  /\sqrt{\text{Var}(Y_{.j})} }{\sum_{i\in[n]} \sum_{j\in[p]} I(\tau_{ij}=0)}
\end{align*}
for a single imputed dataset. Again, we collect the mean and standard deviation of wMSE across all $M$ imputed datasets. 

(iii) Weighted mean bias error (wMBE):
\begin{align*}
    \frac{ \sum_{i\in[n]} \sum_{j\in[p]} I(\tau_{ij}=0)\times (\mathbb{E}[{y_{ij}}^\text{imputed}] - y_{ij}^\text{true}) / \sqrt{\text{Var}(Y_{.j})} }{\sum_{i\in[n]} \sum_{j\in[p]} I(\tau_{ij}=0)}
\end{align*}
across all $M$ imputed datasets, where $\mathbb{E}[{y_{ij}}^\text{imputed}]$ can be estimated by $\frac{\sum_{m\in [M]} y_{ij}^{(m)\text{imputed}}}{M}$.

\begin{figure}[tb]
\makebox[\textwidth][c]{
    \includegraphics[width=0.8\textwidth]{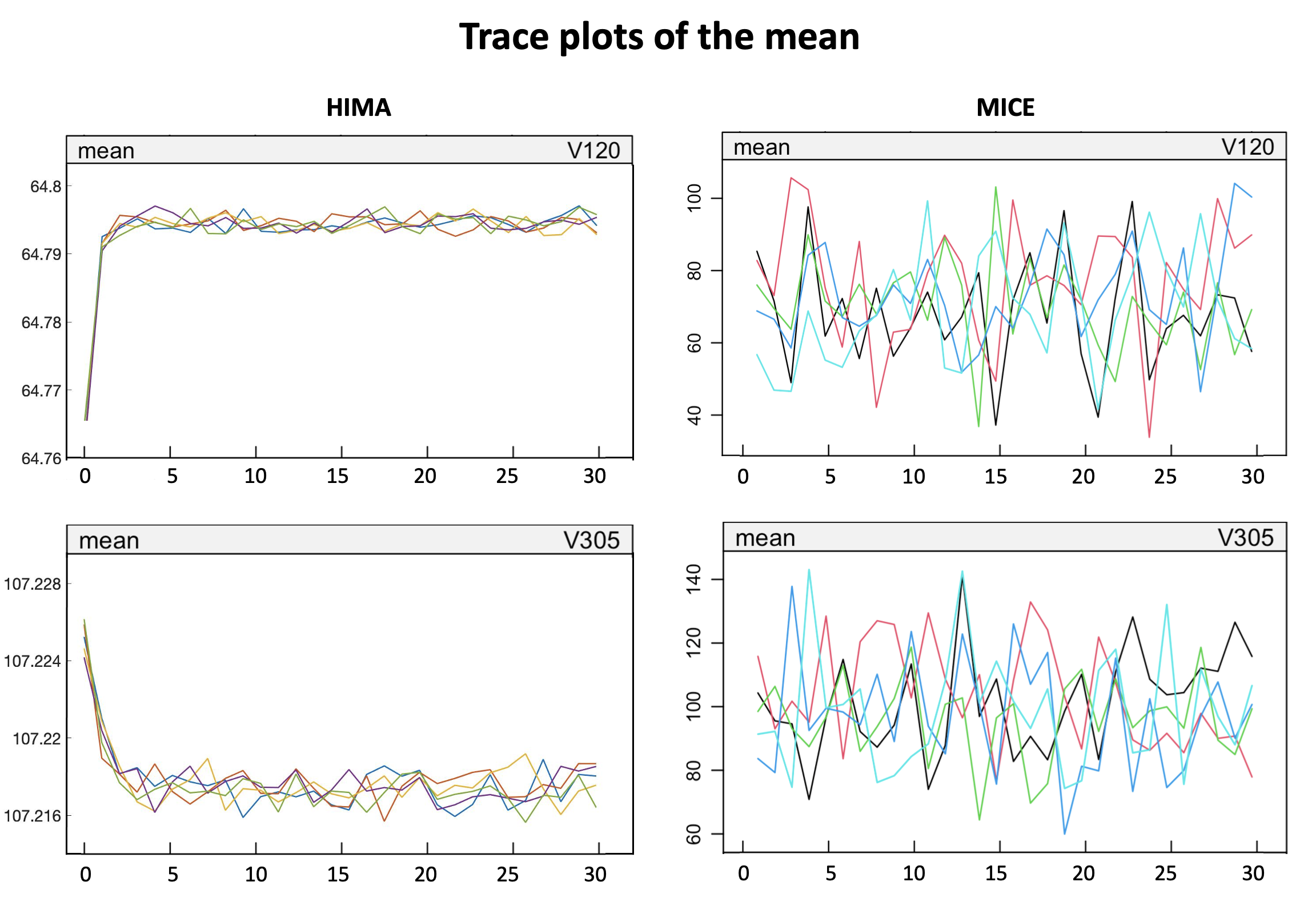}
}
\caption[Trace plots of convergence performance]{\footnotesize \textit{ Trace plots of the means for two randomly selected incomplete voxels $\textit{\#120}$ and $\textit{\#305}$ in the right hippocampus against the number of iterations. The means were estimated iteratively by methods HIMA and MICE, both of which show trends of convergence to stationary, while HIMA converged faster and had reduced dispersion of estimates. Besides, all plots were free of any abnormal trends.  }}
\label{tab:HIMA_traceplots}
\end{figure}

\vspace{2mm}
We select three regions for result demonstration: right insular cortex (Ins), right caudate nucleus (Caud), and right hippocampus (Hippo), which are brain areas frequently associated with information processing in schizophrenia \cite{harrison2004hippocampus, mueller2015abnormalities}.
Based on ICBM, there are respectively $1196, 553,$ and $622$ voxels in the right Ins, right Caud, and right Hippo.  
We applied HIMA on the post-processed data $\boldsymbol{Y}^*_\textbf{Ins}$, $\boldsymbol{Y}^*_\textbf{Caud}$, $\boldsymbol{Y}^*_\textbf{Hippo}$ and compared the imputation performance with frequently used approaches in both simple imputation (e.g., mean-substitution imputation) and multiple imputation (e.g., MICE ).
As mentioned previously, almost all existing MI methods and toolkits for brain imaging data were based on MICE.
We provide the imputation error measure and computational time of these three methods in Table 1. 
In addition, \autoref{tab:HIMA_traceplots} shows the trace plots of estimates over iterations.

\captionsetup{labelformat=empty}
\begin{figure}[!ht]
\makebox[\textwidth][c]{
    \includegraphics[width=\textwidth]{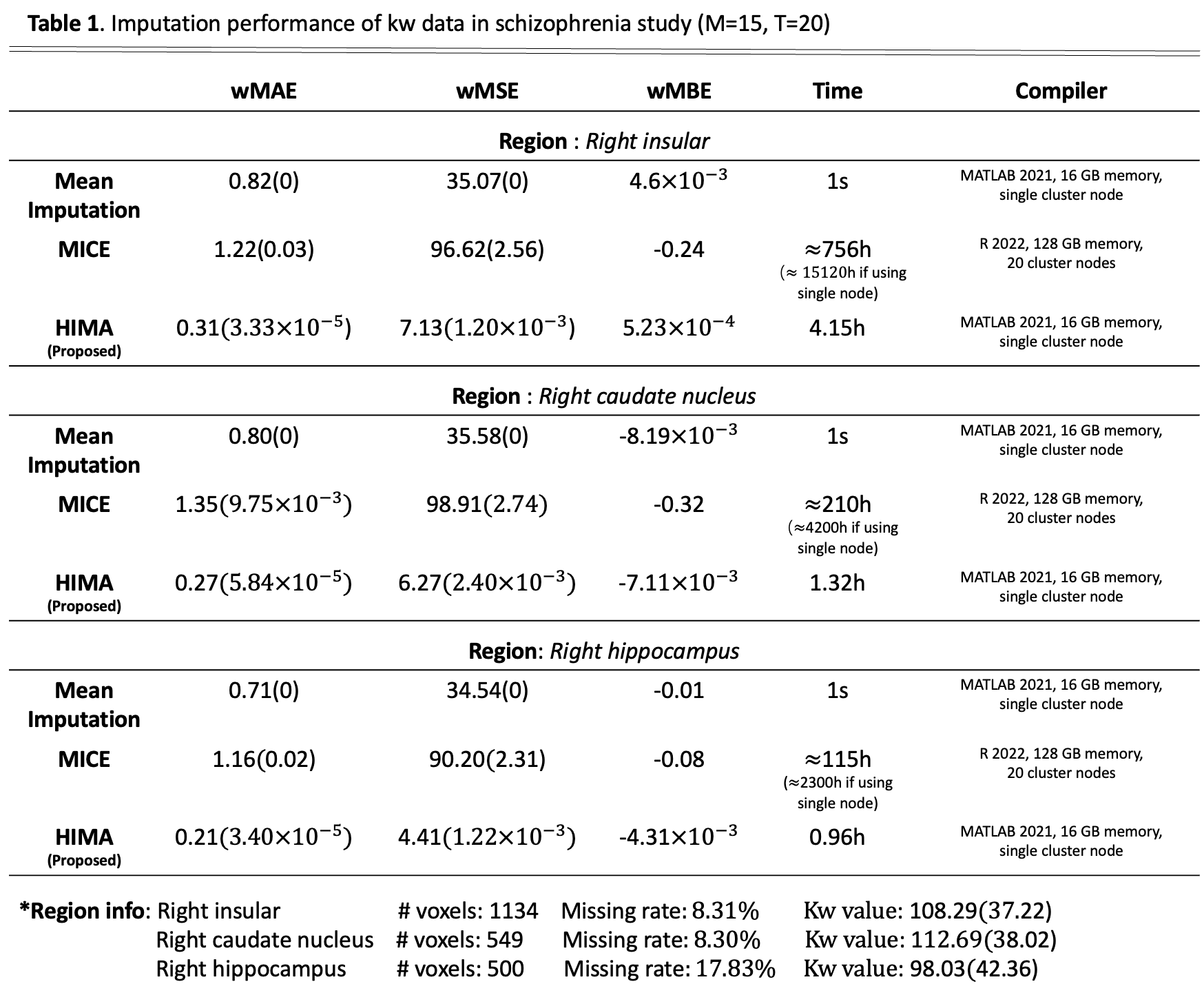}
}

\caption[Imputation performance on semi-synthetic data]{\footnotesize \textit{}}
\label{tab:HIMA_table1}
\end{figure}
\captionsetup{labelformat=default}

\vspace{2mm}
Based on Table 1 and \autoref{tab:HIMA_traceplots}, we asssess the imputation performance from the following three aspects:
\begin{enumerate}
    \item \textbf{Accuracy:} 
    Based on various error metrics (wMAE, wMSE, and wMBE), HIMA shows lower imputation errors in all three selected brain regions, compared to Mean imputation and MICE. Furthermore, the imputed results generated by HIMA exhibit lower variances, indicating high consistency and stability of imputed results. In summary, HIMA demonstrates improved imputation accuracy, as evidenced by the reduction in errors and dispersion.
    
    \item \textbf{Convergence:}
    We created trace plots to visualize the convergence of the estimated mean of Kw values against iteration numbers. 
    \autoref{tab:HIMA_traceplots} showed that both methods had reached a stable posterior distribution after a few iterations, indicating quick convergence to stationarity. With HIMA, voxels converged to stationary estimates with smaller variations, suggesting a higher level of stationarity. Additionally, neither method produced any discernible trends, suggesting sufficient randomness in the estimates across iterations.

    \item \textbf{Computational cost:}
   Mean imputation is the fastest method in terms of running time due to its straightforward operation, as it doesn't require drawing and updating multiple values during each iteration. However, its imputation accuracy is relatively low. In contrast, HIMA employs the principle of MI methods and demonstrates significantly improved computational efficiency compared to MICE, while still reducing imputation errors. For regions with varying sizes and missing rates, HIMA shows over $2,000$ times greater computational efficiency compared to conventional methods, indicating a significant advancement in the computational competency for imputing ultra-high dimensional imaging data.
    
\end{enumerate}

 \subsection{Real data analysis}
In this real data analysis, we apply HIMA on each of the 99 distinct regions without artificially  inserting any missingness this time. Hence, the dataset $\{y_{ij}\}^{r\in [99]}$ to be imputed is the raw MRI data, where its missing entries were mainly caused by image acquisition limitations and susceptibility artifacts. The overall missing rate of the data is $14.6\%$. We first applied the same preprocessing procedures on each $\{y_{ij}\}^{(r)}$: (i) exclude voxels with missing rates higher than $40\%$ to ensure HIMA's stability and accuracy (on average, less than $3\%$ voxels in this real dataset were excluded for each region); (ii) align kernel-smoothed probability density estimate of each subject to eliminate subject-level random effect during imputation. We next applied HIMA on each post-processed data $\{{\boldsymbol{Y}^{(r)}}^*, r\in [99]\}$ with $T=20$ iterations and $M=50$ imputations. Here, we increased the number of imputations to better assess the stability and accuracy of imputed datasets since true information about missing Kw was not available. 

\begin{figure}[!ht]
\makebox[\textwidth][c]{
    \includegraphics[width=1\textwidth]{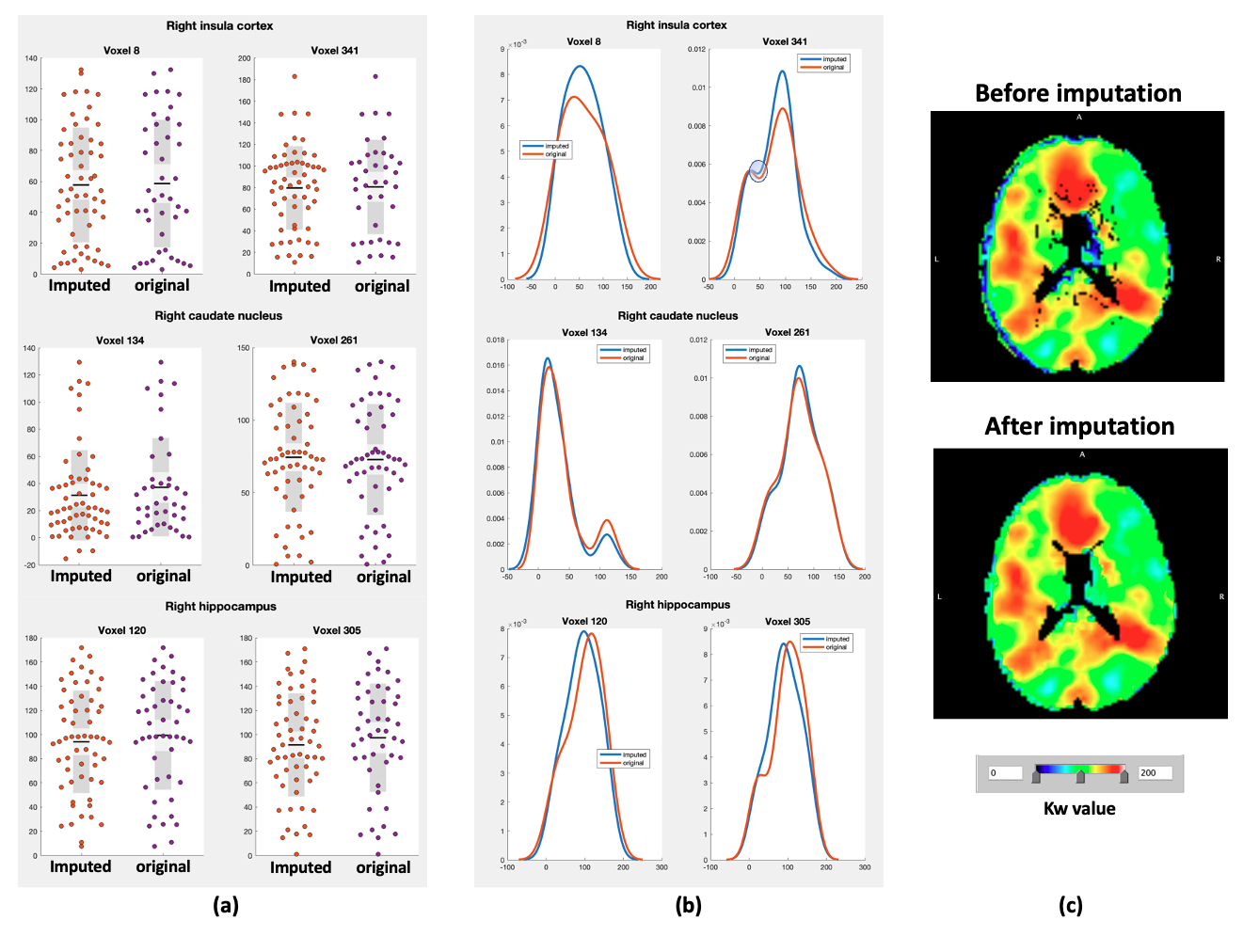}
}
\caption[Imputation results on real schizophrenia data]{\footnotesize \textit{ (a) The shape of the red points (imputed data) matches the shape of the purple ones (observed data). The matching shape indicates the plausibility of the imputed values. (b) The probability density estimate of the observed data is similar to that of imputed data. (c) shows the heatmaps of the whole-brain Kw values from a randomly selected subject before and after imputation using HIMA.  }}
\label{tab:HIMA_real_data}
\end{figure}

We used the same three schizophrenia-associated regions (right Ins, right Caud, and right Hippo) to evaluate the performance of imputation methods. Since the true values of missing data entries are unknown, previous error metrics cannot be evaluated. Instead, we evaluated the results by examining the distributions of the observed and imputed values. 
In \autoref{tab:HIMA_real_data}(a), we plotted the imputed data against the observed data from three randomly selected voxels. The shape of the red points (imputed data) closely matches the shape of the purple dots (observed data). This alignment indicates the plausibility of the imputed values. 
In \autoref{tab:HIMA_real_data}(b), we plotted the probability density estimate of the observed data against the imputed data for the same three voxels. 
The distributions of imputed values are similar to the observed values. In summary, the imputed values in general approximate the observed values, which can facilitate subsequent statistical inference with improved accuracy.  

\section{Discussion}
In this study, we developed a multiple imputation tool, HIMA, specifically designed for analyzing high-throughput multivariate imaging data with missing values. It has been well-studied that simply neglecting missing values or relying on single imputation methods in brain imaging data analysis often leads to suboptimal accuracy in statistical inference \cite{brown2003data, rubin2004multiple, lee2014introduction}. 
In practical applications, however, missing values in neuroimaging data emerge at various spatial locations across different participants, introducing computational challenges for Bayesian-based multivariate MI methods with the MAR assumption. 
Particularly, the high dimensionality of imaging data leads to intractable posterior sampling of large covariance matrices and necessitates computational time of months when using classic multivariate MI tools, such as MICE. To meet the demand, we developed a new computational algorithm with remarkably improved efficiency for implementing the posterior sampling (in minutes).

In addition to the improved computational efficiency, HIMA improves imputation accuracy. Both extensive simulation experiments and real data analysis demonstrated robust and accurate performance of multivariate missing data imputation. HIMA can perform effectively with up to $40\%$ missing observations. Using semi-synthetic data, we showed that the imputed values by HIMA yield less bias in the mean and reduced dispersion when compared to existing methods. In short, HIMA provides a fast and accurate MI solution for multivariate neuroimaging data with varying missing values. The sample codes for HIMA are available at \href{https://github.com/TongLu-bit/HighDim-MultipleImputation-HIMA}{https://github.com/TongLu-bit/HighDim-MultipleImputation-HIMA}.

\hfill \break
\textbf{Declaration of interest}: none.

\hfill \break
\textbf{Acknowledgments}
  This project was in part supported by the National Institutes of Health under Award Numbers 1DP1DA04896801. We would also like to extend our sincere appreciation to Eric Goldwaser and Bhim Adhikari for their efforts in collecting, preprocessing, and providing the imaging data for this research.

\bibliographystyle{apalike}
\bibliography{Reference}

\end{sloppypar}
\end{document}